# Molecular-dynamics simulations of stacking-fault-induced dislocation annihilation in pre-strained ultrathin single-crystalline copper films

Kedarnath Kolluri, M. Rauf Gungor, and Dimitrios Maroudas[*]

Department of Chemical Engineering

University of Massachusetts, Amherst, MA 01003-3110

**Abstract**

We report results of large-scale molecular-dynamics (MD) simulations of dynamic deformation under biaxial tensile strain of pre-strained single-crystalline nanometer-scale-thick face-centered cubic (fcc) copper films. Our results show that stacking faults, which are abundantly present in fcc metals, may play a significant role in the dissociation, cross-slip, and eventual annihilation of dislocations in small-volume structures of fcc metals. The underlying mechanisms are mediated by interactions within and between extended dislocations that lead to annihilation of Shockley partial dislocations or formation of perfect dislocations. Our findings demonstrate dislocation starvation in small-volume structures with ultra-thin film geometry, governed by a mechanism other than dislocation escape to free surfaces, and underline the significant role of geometry in determining the mechanical response of metallic small-volume structures.

---

[*]To whom correspondence should be addressed. E-mail: maroudas@ecs.umass.edu

## I. Introduction

A broad range of modern technologies relies increasingly on the use of nanometer-scale-thick metallic films. Typically, over 500 processes are involved in the development of micro- and nano-scale devices. During these processes, the constituent materials are subjected to various thermo-mechanical environments, which introduce strain in the nano-scale-thick metallic films, causing them to undergo plastic deformation. Such materials response to strain has significant impact on the ductility, hardness, and strength of the metallic films and may lead eventually to their failure; the mechanical response of device interconnects in microelectronics offers typical examples.[1-4] The mechanical behavior of single-crystalline metallic materials undergoing plastic deformation is a direct consequence of the nucleation and motion of dislocations, as well as a variety of dislocation-dislocation interactions. For bulk materials, such mechanisms are well understood and phenomenological models have been developed for their deformation. These phenomenological models, however, cannot be extended to the nanoscale and, more specifically, to describing mechanical response of ultra-thin metallic films.

Unlike bulk metals, metallic nanostructures and other small-volume structures exhibit, for example, ultra-high strength. Examples range from whiskers with small diameters that were reported (as early as almost 6 decades ago) to have much higher strength than bulk metals[5,6] to the uniaxial compression of nanometer-scale pillars of face-centered cubic (fcc) metals that yielded flow strengths comparable to the materials' ideal shear strength.[7-11] In bulk metals subjected to external stress, work hardening is caused by an increase in the dislocation density and by a sharp decrease in plastic flow; the former is due to rapid dislocation multiplication through double cross-slip and Frank-Read-type mechanisms, while the latter is due to interactions between dislocations in multiple slip planes that lead to the formation of sessile dislocations that act as barriers to dislocation motion, as well as due to pile-up of dislocations at obstacles such as grain boundaries and sessile dislocations. In metallic nano-meter-thick films and (in general) small-volume structures, however, it has been argued that depletion of dislocations causes the observed ultra-high strength.[7-13] For example, in compression experiments of Ni single-crystalline nanopillars with a high initial dislocation density, the dislocation density always decreased during the application of strain;[11] subsequently, the nearly defect-free crystal deformed elastically until new dislocations nucleated. In ultra-thin metallic films, a fundamental understanding of the atomistic mechanisms of deformation dynamics is essential for accurate property predictions, better reliability analysis, and design of higher-quality devices. The deformation mechanisms, however, are particularly difficult to study experimentally; for example, in high-strain-rate deformation of single-crystalline fcc metals such as Cu, certain strain relaxation time scales are in the sub-nanosecond range.[14] Atomic-scale and other fine-scale dynamical simulation methods provide ideal means for analyzing the deformation mechanisms in ductile thin films by direct monitoring of the dynamical response of the materials at the atomic/microscopic scale.[15,16]

First-principles density-functional-theory calculations and large-scale molecular-dynamics (MD) simulations have been used extensively toward a fundamental understanding of plastic flow initiation and nucleation of dislocations, as well as investigations of plastic deformation during nanoindentation of metal surfaces.[17,18] These techniques also have been used to explore and analyze dislocation intersections,[19,20] deformation processes at shock loading conditions in metallic bulk[21] and thin-film[22-24] materials, as well as amorphization of fcc crystalline metallic nanowires under strain-rate-limited shock loading at controlled temperatures.[25] In addition, discrete dislocation-dynamics simulations have been used to study size effects in nanopillars of fcc metals; these simulations of fcc nanopillars have shown that dislocation escape through the free surfaces and depletion or inactivation of dislocation sources play a significant role in the plastic response of small-volume structures.[13,26-28] The fundamental mechanisms of dislocation interactions and their possible annihilation in ultra-thin metallic films, however, remain largely unexplored and, therefore, elusive.

In this article, we report results of an atomic-scale analysis of dislocation-dislocation interaction mechanisms based on MD simulations of dynamic deformation of free-standing single-crystalline ultra-thin fcc copper films. The thin-film samples used in our analysis were pre-strained to introduce dislocations, consistently with samples used in experimental studies of the mechanical behavior of small-volume structures that had a high initial dislocation density.[11] Our results indicate that the interactions within and between simple extended dislocations, each of which consists of two Shockley partial dislocations with complementary Burgers vectors separated by a stacking fault, may contribute significantly to dislocation depletion in nanometer-scale fcc metals. We identify different classes of stacking fault-dislocation interactions and present a detailed analysis of dislocation depletion in these films mediated by these mechanisms. Our findings suggest that dislocation escape to free surfaces may not be the primary mechanism of dislocation depletion in small-volume structures of fcc metals with ultra-thin film geometry.

The article is structured as follows. The simulation methods and analysis techniques employed in this study are discussed in Sec. II. The results of our computational analysis, which includes the mechanical behavior of the thin films and the underlying physical mechanisms, are presented in Sec. III. The results of the analysis are discussed in Sec. IV in the context of earlier experimental studies on nanopillars and computational studies of interactions between dislocations and twin boundaries. The limitations of our simulations also are discussed in Sec. IV. Finally, the main conclusions of our study are summarized in Sec. V.

## II. Simulation Methods and Analysis of Simulation Results

In our MD simulations, we employed slab supercells consisting of single-crystalline copper films with periodic boundary conditions applied in the $x$- and $y$-directions in a Cartesian representation; the film planes were oriented normal to the $z$ axis. Vacuum space above the top and below the bottom surfaces of the films was used in order to generate free surfaces and mimic free-standing thin films. The Cartesian $x$, $y$, and $z$ axes were taken along the $[\bar{1}10]$, $[\bar{1}\bar{1}2]$, and $[111]$ crystallographic directions, respectively, with the film plane corresponding to the $(111)$ crystallographic plane. For the results reported here, we employed two different simulation cells: one that contained 1,520,640 Cu atoms (with an edge size of 61.35, 63.76, and 4.38 nm in the $x$-, $y$-, and $z$-direction, respectively) and another one that contained 1,548,797 Cu atoms (with an edge size of 40.90, 38.96, and 11.27 nm in the $x$-, $y$-, and $z$-direction, respectively); in both cases, the reported cell dimensions correspond to the unstrained state of the metallic thin film.

The interatomic interactions were modeled using an embedded-atom-method (EAM) potential, parameterized for copper.[29] This potential was validated by comparison of its predictions with those of *ab initio* calculations.[30] For our MD simulations of thin-film dynamic deformation under biaxial tensile strain, we used the public-domain computer software LAMMPS.[31] In our simulations, the temperature was kept constant at $T = 100$ K by explicitly rescaling the atomic velocities at each time step. In the MD simulations of thin-film dynamic deformation under biaxial tensile strain, all the films used were pre-treated before conducting the MD simulations of dynamical straining in order to generate the initial dislocation distributions. Specifically, the thin films were pre-strained in biaxial tension to a specified strain level at a constant true strain rate of $2.01 \times 10^{11}$ s$^{-1}$ (equally in each lateral direction). Under this mode of loading, three {111} crystallographic planes are activated; the Schmidt factor for all possible slip directions in the (111) plane, which is also the orientation of the thin film surface, is negligibly small. After the desired strain level was reached, isothermal-isostrain MD simulations were performed in order to equilibrate the thin-film structure. We analyzed the mechanical response of three thin-film samples; two of these, samples 1 and 2, had initial (unstrained-state) thicknesses of 4.38 nm and 11.27 nm, respectively. The third one, sample 3, was prepared by compressing sample 1 after its tensile straining back to its original lateral dimensions at the same constant high strain rate and then equilibrating the film's structure by isothermal-isostrain MD simulation. In their pre-treatment, the thin-film samples discussed in this article were subjected to biaxial tensile strain at a level of 8%. For pre-treatment under biaxial tensile strain at strain levels over the range from 6% to 10%, the resulting microstructure of the equilibrated thin films remained qualitatively identical. The final thicknesses of samples 1, 2, and 3 were 3.84 nm, 9.76 nm, and 4.43 nm, respectively.

These samples contained predominantly simple extended dislocations that either extended across the film thickness or were threading dislocation half loops emanating from the film surfaces. The initial dislocation density, $\rho$, of these films ranged from 1.0 × $10^{17}$ m$^{-2}$ to 3.2 × $10^{17}$ m$^{-2}$; $\rho = l/V$ where $l$ is the total length of all the dislocations in the film and $V$ is the film's volume. In samples 1 and 3, many extended dislocations spanned across the film thickness, while only a few threading dislocation half loops were present. In the thicker film, however, the population of threading dislocation half loops was comparable with that of the extended dislocations that spanned across the film thickness. We note that this initial film microstructure is similar to that of the Ni nanopillars used in mechanical annealing experiments;[11] in those experiments, similar dislocation types were observed, i.e., most of them extended across the thickness of the nanopillar and a few of them were threading dislocation half loops at the surface of the nanopillar, and the dislocation density was ~ $10^{15}$ m$^{-2}$. After this film pre-treatment stage, the final steady-state values of the von Mises shear stress, $\sigma_{vM}$, for samples 1, 2, and 3 were 530, 460, and 105 MPa, respectively; $\sigma_{vM}$ is defined as the second invariant of the stress tensor and is given by $\sigma_{vM} = \frac{1}{\sqrt{6}} \sqrt{(\sigma_{xx} - \sigma_{yy})^2 + (\sigma_{yy} - \sigma_{zz})^2 + (\sigma_{zz} - \sigma_{xx})^2 + 6(\sigma_{xy}^2 + \sigma_{yz}^2 + \sigma_{zx}^2)}$, where $\sigma_{\alpha\beta}$ ($\alpha, \beta = x, y,$ or $z$) are the elements of the stress tensor. Although the stresses in the thin films were significant, most of the stress was due to elastic straining. In the thin films studied here, the shear strain, measured as the second invariant of the local atomic strain tensor,[32] varied between 0.7% and 1.59%; this indicates that in these pre-treated thin films the applied biaxial strain had relaxed significantly. Additionally, the pre-treatment equilibration stage was continued until ensemble averages of the stress and potential energy in the thin films reached constant (i.e., steady) values. Following pre-treatment, the samples were strained biaxially in tension at a constant strain rate of 7 × $10^8$ s$^{-1}$ (again equally in each lateral direction) and the evolution of the film structure was monitored in detail.

For monitoring the structural response of the thin film to the applied biaxial strain, we employed common neighbor analysis (CNA) in order to identify atoms in locally perfect hexagonal close-packed (hcp) and fcc lattice arrangements.[33] We used the public-domain computer software ATOMEYE [32] in order to visualize sequences of MD-generated atomic configurations. Atoms with 12 nearest neighbors (coordination number for perfect fcc and hcp lattices) that are not in locally perfect fcc or hcp lattice arrangements are located in the dislocation cores; in our analysis, we have identified these "non-coordination lattice defects" as a means of dislocation visualization. The dislocation densities computed in our simulations, on the order of $10^{16}$ - $10^{17}$ m$^{-2}$, are one to four orders of magnitude higher than those observed typically in experiments; such high dislocation densities are typical of MD simulations.

The stress in the material was calculated using the Virial formula.[34] We implemented two methods for the computation of the film's volume (involved in both the stress and the dislocation density computation). In the first method, we constructed on the thin-film surface a grid of multiple elements of known lateral dimensions; we computed the thickness of each such element as the distance, in $z$, between the two atoms on the opposite surfaces of the film that were the farthest located from its center plane on each surface within the element. The total volume of the film was then computed as the sum of the element volumes. In the second method, we estimated the thickness of the film as the distance, in $z$, between the two atoms on each of the opposite film surfaces that were the farthest located from the center plane of the film. The first method underestimates the total film volume, while the second method overestimates it. The stress values reported in the stress-strain curve plots of this article were computed using the second method. Irrespective of the method used for film volume computation, the corresponding qualitative features of the stress evolution in dynamic deformation remained identical except at a stage close to failure; at this stage, the volume computed according to the first method excludes (incorrectly) the pits formed on the surface. The difference in the stress values computed by the two methods were ~100-200 MPa; the first method yielded consistently higher stress values.

## III. Mechanical Behavior and Underlying Physical Mechanisms

Figure 1(a) shows the evolution of the von Mises shear flow stress, $\sigma_{vM}$, dislocation density, $\rho$, and the stacking fault area, $A_{sf}$, in sample 1 (initial $\rho = 1.4 \times 10^{17}$ m$^{-2}$) during dynamical biaxial straining, where the applied strain level, $\varepsilon$, increases linearly with time. The stacking fault area is defined as $A_{sf} = (N_{hcp}/2) \times (1/3) \times (3\sqrt{3}a^2/4)$, where the first factor is the number of hcp atoms on one of the two planes of the stacking fault, the second factor is the number of atoms per hexagonal unit of the stacking-fault lattice plane, and the third factor is the area of each such hexagonal unit. The decrease in $\rho$ and $A_{sf}$ until they each reach a minimum value indicates annihilation of dislocations. Subsequently, new dislocations nucleate, leading to an increase in $\rho$ and $A_{sf}$. From the shear stress-strain curve, it is evident that the thin film does not undergo stage-II hardening, which is typical of bulk metals. The maximum $\sigma_{vM}$ computed in our simulations varied between 1.56 and 1.74 GPa. These stress levels are 39.9%-44.5% of the ideal simple shear strength of our copper model (3.91 GPa according to the interatomic potential employed in our MD simulations, which compares well with first-principles calculations[30]); we also note that these stress levels are consistent with experimental measurements on Au nanopillars (indicating that the nanopillars reached 44% of the material's theoretically predicted ideal pure shear strength[12]).

The mechanical response of the thin film to the applied biaxial strain exhibits three different stages; these stages are indicated in Fig. 1(a) and are termed as Stage I, Stage II, and Stage III. Figure 1(b) shows three-dimensional views of the MD simulation cell at various states, which are also marked in Fig. 1(a). In Stage I, $\sigma_{vM}$ increases rapidly to about 1.1 GPa. The glide of most dislocations is activated in this stage. A few partial dislocations retract into the extended dislocations that they constitute and, in the process, they unzip the stacking faults that they bound. Such dislocation glide induces the interaction of the two Shockley partial dislocation components of a simple extended dislocation, resulting in a reduction of the stacking-fault area. The reduced stacking-fault area increases the mobility of the dislocations. In addition to the reduction in stacking-fault area, a few dislocations also annihilate in Stage I, primarily due to interactions between extended dislocations and, to a far lesser extent, due to interactions between partials and anti-partials that bind a few stacking faults. Dislocation annihilation due to interactions between partial and anti-partial dislocations occurs possibly because such dislocations are unstable. In Stage II, for 1.5% < ε < 4.5% [Fig. 1(a)], $\rho$ and $A_{sf}$ decrease to less than one third of their initial values; $\sigma_{vM}$, however, remains nearly constant. Stage II is dominated by dislocation glide and, more importantly, by interactions within and between extended dislocations. The intersection of the gliding dislocations with the stacking faults of the extended dislocations causes cross-slip of the intersecting dislocations and break-up of the obstacle stacking faults. At least one of the dislocations produced by the cross-slip reaction glides in the plane of the obstacle extended dislocation and interacts with its Shockley partials to either form a perfect dislocation or annihilate. Such stacking fault-aided cross-slip mechanisms cause further reduction in $\rho$ and $A_{sf}$ [Fig. 1(b), state (iii) at the end of Stage II]. In Stage III, the dislocation dynamics is similar to that observed in Stage II. By the end of Stage III, $\rho$ and $A_{sf}$ reach their minimum values; most dislocations that survive Stage III are perfect dislocations. Unlike Stage II, in Stage III, $\sigma_{vM}$ increases with increasing $\varepsilon$ due to reduced plastic flow (dislocation starvation) in the thin film. At the end of Stage III, after a considerable build-up in shear stress, new dislocations start nucleating at the thin-film surface, leading to a sharp decrease in the thin-film stress. In sample 1 (Fig. 1), the propagation of the newly nucleated dislocations leads quickly to film failure; we note that, due to film failure, we do not observe successive strain burst events similar to those observed in experiments on metallic nanopillars.[9-13]

Sample 3 also responds to applied biaxial strain in a similar fashion. The evolution of $\sigma_{vM}$, $\rho$, and $A_{sf}$ in samples 2 and 3 is shown in Fig. 2(a) and Fig. 2(b), respectively. In samples 1 and 3, the stress increases rapidly and then remains constant until significant dislocation depletion and annihilation of stacking faults occur. Reduction in plastic flow due to dislocation depletion leads to build-up of stress in the thin films and, eventually, to nucleation of new dislocations. As dislocation density in sample 3 reaches a minimum faster (due to much higher initial dislocation density than in sample 1), the regime of constant stress is smaller in sample 3 than in sample 1. Samples 1 and 3 are considerably thin; in these samples, the nucleation and propagation of new dislocations lead quickly to thin-film failure. In sample 2, however, the newly nucleated dislocations propagate without causing failure; instead, they lead to successive cycles of dislocation dissociation and annihilation followed by dislocation nucleation events. Consequently, the strain bursts that correspond to nucleation of new dislocations and the associated sharp decrease in the thin-film stress are identified clearly in Fig. 2(a); here, it is worth noting that these strain bursts are similar to those observed in the nanopillar experiments.[9-13] In Fig. 2(a), green shaded zones correspond to deformation stages associated with plateaus and local minima in the stress curve; these stages are characterized by a decrease in the stacking fault area and dislocation density, indicating that these deformation regimes of constant and decreasing stress are accompanied by dislocation dissociation and annihilation. On the other hand, red shaded regions in Fig. 2(a) are associated with local maxima in the stress curve and are characterized by an increase in the dislocation density and the stacking-fault area, indicating nucleation of new dislocations and their propagation in the thin film. Videos of the full dynamic deformation simulations of samples 1 and 2 can be found in the *Supplementary Information*.[44]

Through detailed analysis of our simulation results, we have identified the various interactions between the gliding dislocations and the stacking faults that lie in their path. These interactions can be classified as (I) reactions between the stacking fault of an obstacle extended dislocation and a mobile perfect dislocation: the Burgers vector, **b**, of the mobile perfect dislocation is such that the dislocation (a) can or (b) cannot cross-slip into the plane of the obstacle extended dislocation without dissociation; and (II) reactions between a partial dislocation and the stacking fault of an obstacle extended dislocation.

Figure 3(a) shows atomic configurations along the pathway of a reaction under category I(a). An extended dislocation is shown to glide in the $(1\bar{1}1)$ plane, consisting of a leading and a trailing Shockley partial with **b** = $(a/6)[21\bar{1}]$ and $(a/6)[121]$, respectively, where $a$ is the fcc lattice parameter. A second extended dislocation lies in the $(1\bar{1}\bar{1})$ plane and is bounded by Shockley partials with **b** = $(a/6)[\bar{1}1\bar{2}]$ and $(a/6)[211]$. The stacking fault of the second extended dislocation acts as an obstacle to the glide of the first one. When the leading partial of the mobile extended dislocation approaches the obstacle stacking fault, it "waits" for the trailing partial to "catch up." This causes the combination of the two partials to form a perfect dislocation with **b** = $(a/2)[110]$. The resulting perfect dislocation glides to and dissociates on the obstacle stacking fault; from this dissociation, two Shockley partial dislocations are formed, with **b** = $(a/6)[12\bar{1}]$ and $(a/6)[211]$, which glide on the $(1\bar{1}\bar{1})$ crystallographic plane. The corresponding combination and dissociation reactions are given by

$$(a/6)[21\bar{1}](1\bar{1}1) + (a/6)[121](1\bar{1}1) \rightarrow (a/2)[110](1\bar{1}1) \text{ and}$$

$$(a/2)[110](1\bar{1}1) \rightarrow (a/6)[211](1\bar{1}\bar{1}) + (a/6)[12\bar{1}](1\bar{1}\bar{1}).$$

The glide of the new Shockley partials unzips the obstacle stacking fault and leads to the formation of two extended dislocations in the same plane; the partials of each of the two extended dislocations can combine to either form a perfect dislocation or annihilate if they have opposite Burgers vectors. In our simulations, combination of at least one pair of partials to form a perfect dislocation is observed frequently.

Reactions under category I(b) also were commonly observed in our MD simulations. Figure 3(b) shows atomic configurations along the pathway for one such reaction. The stacking fault of the obstacle extended dislocation depicted in Fig. 3(b) lies on the $(\bar{1}11)$ plane. A perfect dislocation with **b** = $(a/2)[011]$ is shown to glide on the $(1\bar{1}\bar{1})$ crystallographic plane toward the obstacle stacking fault. Upon intersecting with the obstacle stacking fault, the perfect dislocation dissociates according to the dissociation reaction

$$(a/2)[01\bar{1}](1\bar{1}\bar{1}) \rightarrow (a/6)[11\bar{2}](1\bar{1}\bar{1}) + (a/6)[\bar{3}10] + (a/6)[2\bar{1}\bar{1}](\bar{1}1\bar{1})$$

to form two Shockley partials, one in the plane of the gliding perfect dislocation and another one in the plane of the obstacle stacking fault, and a sessile stair-rod dislocation with $\mathbf{b} = (a/6)[\bar{3}10]$. Subsequently, the produced stair-rod dislocation dissociates according to the reaction

$$(a/6)[\bar{3}10] \rightarrow (a/6)[\bar{1}2\bar{1}](1\bar{1}\bar{1}) + (a/6)[\bar{2}\bar{1}\bar{1}](\bar{1}1\bar{1}).$$

The outcome of the two dissociation reactions is the formation of two extended dislocations in the final state. In effect, upon interacting with the obstacle extended dislocation, a perfect dislocation dissociates into its corresponding Shockley partials and splits the obstacle extended dislocations into two extended dislocations. In our simulations, we observe many reactions classified under category I(a) and I(b), which are similar to the Friedel-Escaig cross-slip mechanism.[35]

Less frequently, we observe the dissociation of a Shockley partial at the stacking fault of an obstacle extended dislocation. Figure 3(c) shows atomic configurations along the pathway of one such reaction. The obstacle and the gliding extended dislocations lie on the $(1\bar{1}\bar{1})$ and $(\bar{1}1\bar{1})$ crystallographic planes, respectively. The leading and the trailing partials of the gliding extended dislocation have $\mathbf{b} = (a/6)[1\bar{1}\bar{2}]$ and $(a/6)[\bar{1}2\bar{1}]$, respectively. Initially, the trailing partial of the gliding extended dislocation is entangled with other dislocations and it is, therefore, unable to glide toward the leading partial, combine with it, and form a perfect dislocation. Upon contact with the obstacle stacking fault, the leading partial dissociates to generate a Shockley partial in the plane of the obstacle stacking fault and a sessile stair-rod dislocation according to the reaction

$$(a/6)[1\bar{1}\bar{2}](1\bar{1}1) \rightarrow (a/6)[\bar{1}\,\bar{1}\bar{2}](1\bar{1}\bar{1}) + (a/3)[100].$$

The trailing partial of the gliding extended dislocation, which is now separated from the entanglement, glides to react with the stair-rod dislocation according to the reaction

$$(a/6)[\bar{1}\bar{2}\bar{1}](1\bar{1}1) + (a/3)[100] \rightarrow (a/6)[1\bar{2}\bar{1}](1\bar{1}\bar{1}).$$

Effectively, this reaction leads to annihilation of the gliding extended dislocation and break-up of the obstacle dislocation into two extended dislocations. This reaction mechanism is similar to the Fleischer cross-slip mechanism.[36] As shown in Fig. 3(c), one of these two dislocations is annihilated.

In the thicker film (sample 2), we also observed dislocation nucleation mechanisms at the intersection of the thin-film surface and the stacking fault of a threading dislocation. Such a newly nucleated threading dislocation half loop grows to unzip an already existing stacking fault and to meet, eventually, the Shockley partial of the initially present threading dislocation; this results either in formation of a perfect dislocation or in complete annihilation of the threading dislocation half loop. Such dislocation nucleation and interaction mechanisms have been observed in simulations of pristine gold nanopillars[37] and bulk aluminum,[38] as well as in simulations of structural relaxation of copper thin films after shock loading.[39] Additionally, our simulations reveal interaction mechanisms where (i) the leading partial transmits through the obstacle extended dislocation irrespective of the glide of the trailing partial and (ii) an obstacle extended dislocation deflects the gliding perfect dislocation to another slip plane without any dissociation or transmission. The former of these interactions belongs to category II, while the latter one belongs to category I(b). These interaction mechanisms, however, were observed only rarely.

IV. **Discussion**

A stage of constant or decreasing stress during dynamic deformation, known as the "easy glide" stage, also has been observed in the compression experiments of Ni and Au nanopillars.[7-12] In the nanopillar experiments, this regime was followed by a regime of increasing stress. The stress increase in these nanopillars has been attributed to reduced plastic flow due to a reduction in the dislocation density. In the case of Ni nanopillars which had a very high initial dislocation density ($\sim 10^{15}$ m$^{-2}$), the experiments showed that the dislocation density at the end of the "easy glide" stage was always far lower than that at the start; in some cases, the nanopillars were dislocation free before a steep increase in the stress.[11] Even though the dislocation annihilation mechanisms that we have observed in our simulations are different from those discussed in the experiments of nanopillar compression, the mechanical response of the ultra-thin films in our simulations is similar to that observed in the experiments on nanopillars of Ni and Au crystals; during the application of strain, the nanopillar strength reaches levels that are comparable to those of the ultra-thin film strength in our simulations, i.e., a substantial fraction (~ 40%) of the material's ideal shear strength.

On the other hand, the mechanical behavior of nanopillars of molybdenum, a body-centered cubic (bcc) metal, was found to be similar to that of the bulk material and no size effects were reported;[12] bcc Mo pillars reached only 7% of bulk molybdenum's ideal strength, while pillars of fcc Au reached 44% of the bulk crystal's ideal strength.[12] Shockley partial dislocations and perfect dislocations are prevalent in fcc and bcc metals, respectively. The glide of perfect dislocations does not leave behind any stacking faults and these dislocations can easily cross-slip without dissociation; as a result, they are not constrained to glide in the same slip plane. Shockley partials, however, can not cross-slip as easily as the perfect dislocations can without dissociation; therefore, they are constrained to their glide plane and leave stacking faults behind as they glide. These differences in the dislocation dynamics between fcc and bcc metals are likely to cause the differences in their mechanical behavior. Our finding that stacking faults play a significant role in the annihilation of dislocations in ultra-thin films of fcc metals may provide a possible explanation for the experimentally observed differences in the mechanical behavior of fcc and bcc metallic nanopillars.

Friedel-Escaig- and Fleischer-type cross-slip mechanisms similar to those identified in our analysis also have been reported in recent studies of dislocation interactions with twin boundaries.[40-43] Particularly, twin-boundary-mediated dislocation dissociation reactions (also termed slip transfer reactions) were shown to be the rate controlling mechanisms of plastic flow in nano-twinned materials and their high ductility was attributed to the structural evolution caused by such mechanisms.[43] These twin-mediated mechanisms are similar to those in categories I and II in our analysis; for the twin-mediated mechanisms similar to those in categories I(a) and II, the activation barrier was reported to be ~0.49 eV, whereas for mechanisms similar to those in category I(b), it was reported to be ~0.67 eV.[43] With stacking faults being the obstacles to dislocation motion in our simulations (as opposed to twin boundaries), the activation energies for similar mechanisms at similar high-stress conditions are expected to be lower than those reported in the twin-boundary-dislocation interaction studies.

Although our simulations have addressed relatively simple model systems, they provide insights into mechanisms that cannot be observed directly in experiments; as a result, the simulation predictions are valuable in designing experimental and computational protocols in order to understand further the role of dislocation interactions in ultra-thin metallic films. There are, however, serious limitations regarding length and time scales that can be accessed directly by MD simulation; in spite of significant recent advances in high-performance computing, the associated computational costs are very high. Consequently, we have not yet analyzed systematically size effects on the mechanical behavior; clearly, the three different film thicknesses used in our simulations are not sufficient to draw any definitive conclusions on size effects. Additionally, MD simulations are limited to high strain rates that are typically orders of magnitude greater than those applied in dynamic deformation experiments. Hence, it is imperative that the characteristic stresses and time scales associated with the deformation-rate-limiting dynamical processes be computed and compared with the corresponding experimentally measured material parameters.

In our MD simulations, the highest stresses computed were far below the ideal shear strength of the material; the maximum shear stresses observed were up to 45% of the ideal shear stress. Moreover, the maximum dislocation velocities computed in our simulations ranged between 400 m/s and 500 m/s, which is well below the sound velocity in the material (3703 m/s for this potential under unstrained conditions). These comparisons suggest that the high strain rates applied in our MD simulations of dynamic deformation do not affect the underlying dislocation dynamics. The initial dislocation density in our study is on the order of $10^{17}$ m$^{-2}$, which is considerably higher than the initial dislocation density of $10^{15}$ m$^{-2}$ in the submicron-diameter Ni pillar compression experiments.[11] This high dislocation density is mainly due to the nanometer-scale film thickness and the thin-film geometry, where dislocation escape to the free surfaces is not likely since the dislocation lines terminate at and are confined between the parallel free surfaces of the thin film. Furthermore, high dislocation densities in MD simulations are inevitable; for example, the dislocation density in the thin films used in our simulations with only a single dislocation extended across the film thickness is ~$7.9 \times 10^{14}$ m$^{-2}$. The reported mechanisms of dislocation depletion, however, are not artificially induced due to the high initial dislocation density since these mechanisms are similar to those identified in recent studies of dislocation interactions with twin boundaries in samples with lower dislocation densities.[40-43] Finally, the initially deformed (pre-strained) structures preceding the dynamical loading are not stress free after the thermal equilibration procedure. The resulting residual elastic strain in our thin films at the beginning of the dynamic deformation simulations represents more realistically the residual strains in the metallic thin films in complex and heterogeneous device structures.

## V. Summary

We have analyzed the mechanical behavior of nanometer-scale-thick fcc Cu films based on large-scale MD simulations of dynamic deformation under biaxial tensile strain. Our analysis identified the physical mechanisms of defect dynamics that govern this mechanical behavior. Specifically, the analysis revealed that stacking faults are likely to provide an important source of dissociation, cross-slip, and annihilation of dislocations in small-volume structures of fcc metals. Our findings suggest that in small-volume structures with ultra-thin film geometry, dislocation escape to free surfaces may not be the primary mechanism of dislocation starvation.

The mechanisms analyzed in this study lead to dislocation annihilation or formation of perfect dislocations. In geometries such as those in nanopillars, these perfect dislocations can readily escape to the free surfaces. Our findings lend support to the dislocation starvation phenomenon and expand our current understanding of dislocation depletion mechanisms in fcc metals. First-principles calculations of the activation energy barriers of stacking fault-mediated dislocation dissociation pathways and their comparison with measurements from carefully designed experiments can shed more light on the role of stacking faults in the mechanical behavior of nanometer-scale structures of fcc metals.

**Acknowledgements**

This work was supported by the Office of Basic Energy Sciences, U.S. Department of Energy through Grant No. DE-FG02-07ER46407. Supercomputing facilities were made available by the National Science Foundation through TeraGrid resources provided by NCSA.

**Figure Captions**

**Figure 1.** (color online) (a) Evolution (i.e., dependence on strain during dynamic deformation) of the shear stress (black), dislocation density (red), and stacking-fault area (blue) computed from molecular-dynamics simulation of biaxial tensile straining of a 3.84-nm-thick Cu film. (b) Three-dimensional view of the thin film at the various different states, (i), (ii), (iii), and (iv), marked in (a); videos of the full simulation are provided in the *Supplementary Information*.[44] Atoms in locally perfect hcp arrangements are colored blue; two consecutive layers of hcp atoms form a stacking fault. Atoms located inside dislocation cores also are colored. Atoms in locally perfect fcc arrangements and the surface atoms are not shown for clarity.

**Figure 2.** (color online) Evolution (i.e., dependence on strain during dynamic deformation) of the shear stress (black), dislocation density (red), and stacking-fault area (blue) computed from molecular-dynamics simulations of biaxial tensile straining of Cu films with initial thickness (a) 9.76 nm (sample 2) and (b) 4.43 nm (sample 3); for (a), videos of the full simulation are provided in the *Supplementary Information*.[44] In (a), deformation stages governed by dislocation nucleation and propagation and by dislocation dissociation and annihilation are depicted as red and green shaded zones, respectively.

**Figure 3.** (color online) Interactions between gliding dislocations and stacking faults in a Cu ultra-thin film (sample 1) under biaxial tensile strain. Atomic configurations are shown along pathways of dislocation-stacking-fault reaction mechanisms; the structural evolution is from left to right. (a) A gliding extended dislocation effectively cross-slips

into the plane of an obstacle stacking fault. The extended dislocation dissociates to form another extended dislocation on the obstacle stacking-fault plane. (b) A gliding perfect dislocation intersects with an obstacle stacking fault and dissociates effectively into two extended dislocations, one of which propagates on the obstacle stacking-fault plane, while the other one breaks through the obstacle stacking fault and glides on the same {111} plane as that of the initial perfect dislocation. (c) A partial dislocation dissociates upon intersecting with an obstacle stacking fault. The product dislocations then glide in the plane of the obstacle stacking fault and unzip the stacking fault. Atoms in locally perfect hcp arrangements are colored blue; two consecutive layers of hcp atoms form a stacking fault. Atoms located inside dislocation cores also are colored. Atoms in locally perfect fcc arrangements and the surface atoms are not shown for clarity.

**Figure. 1**

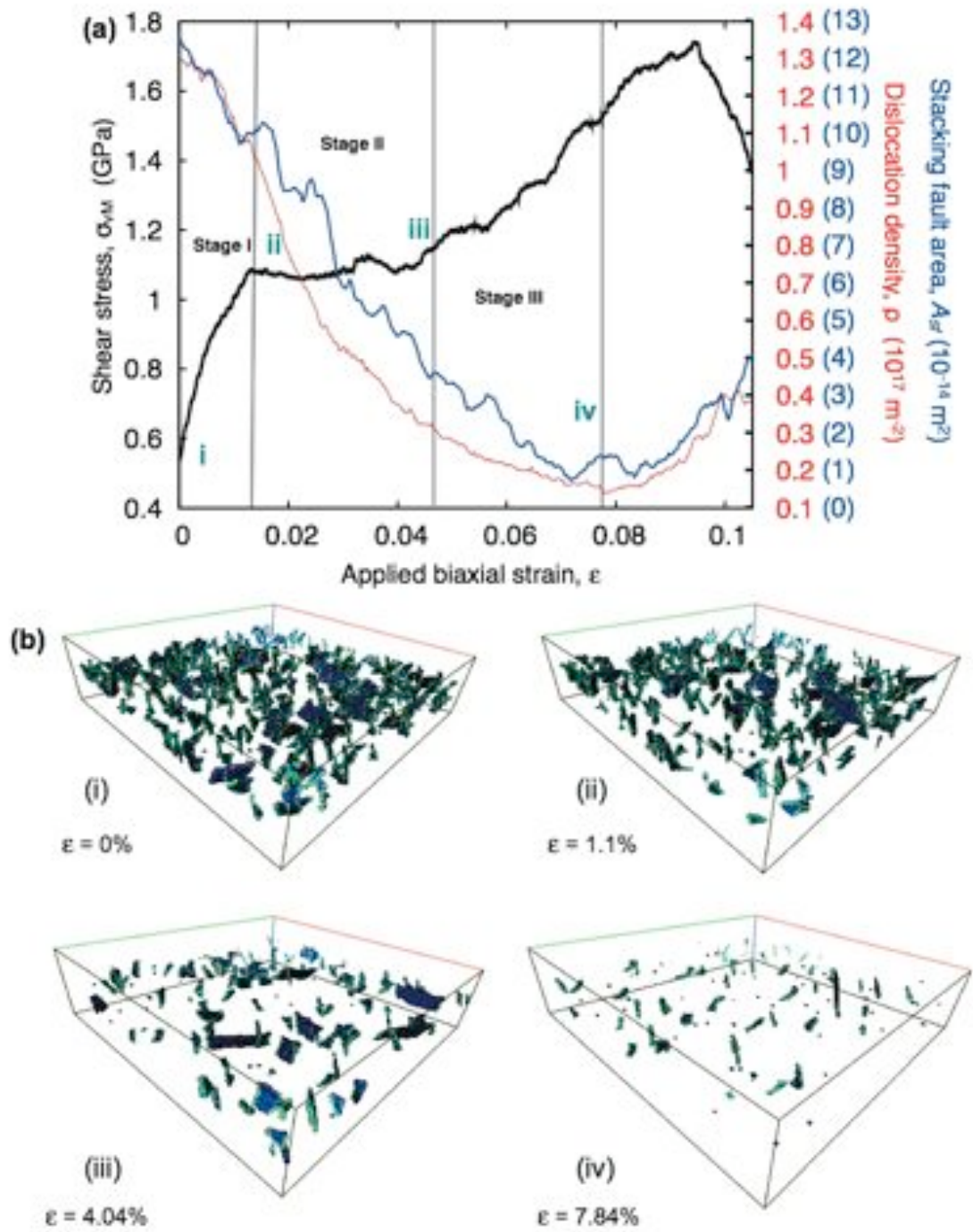

**Figure. 2**

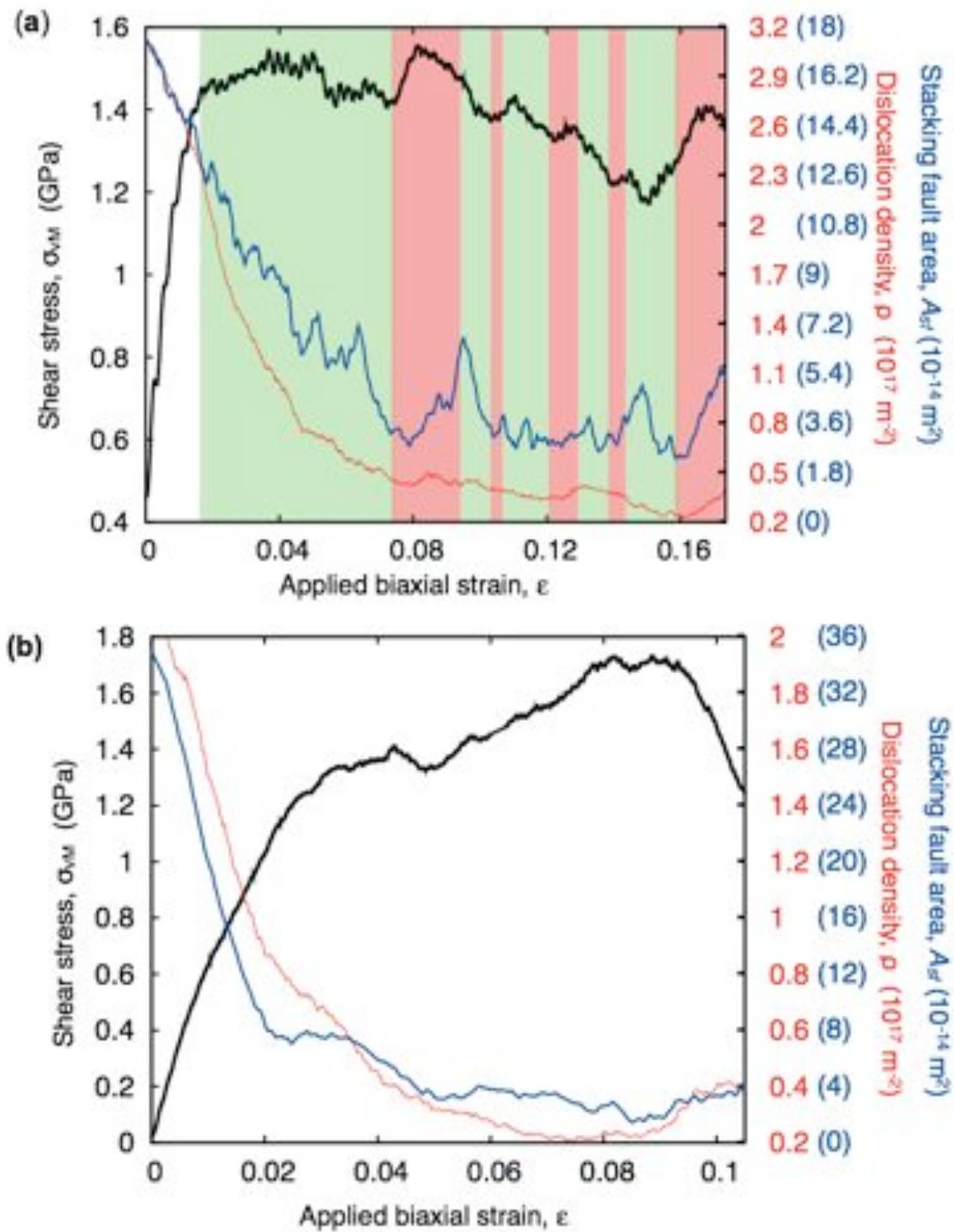

**Figure. 3**

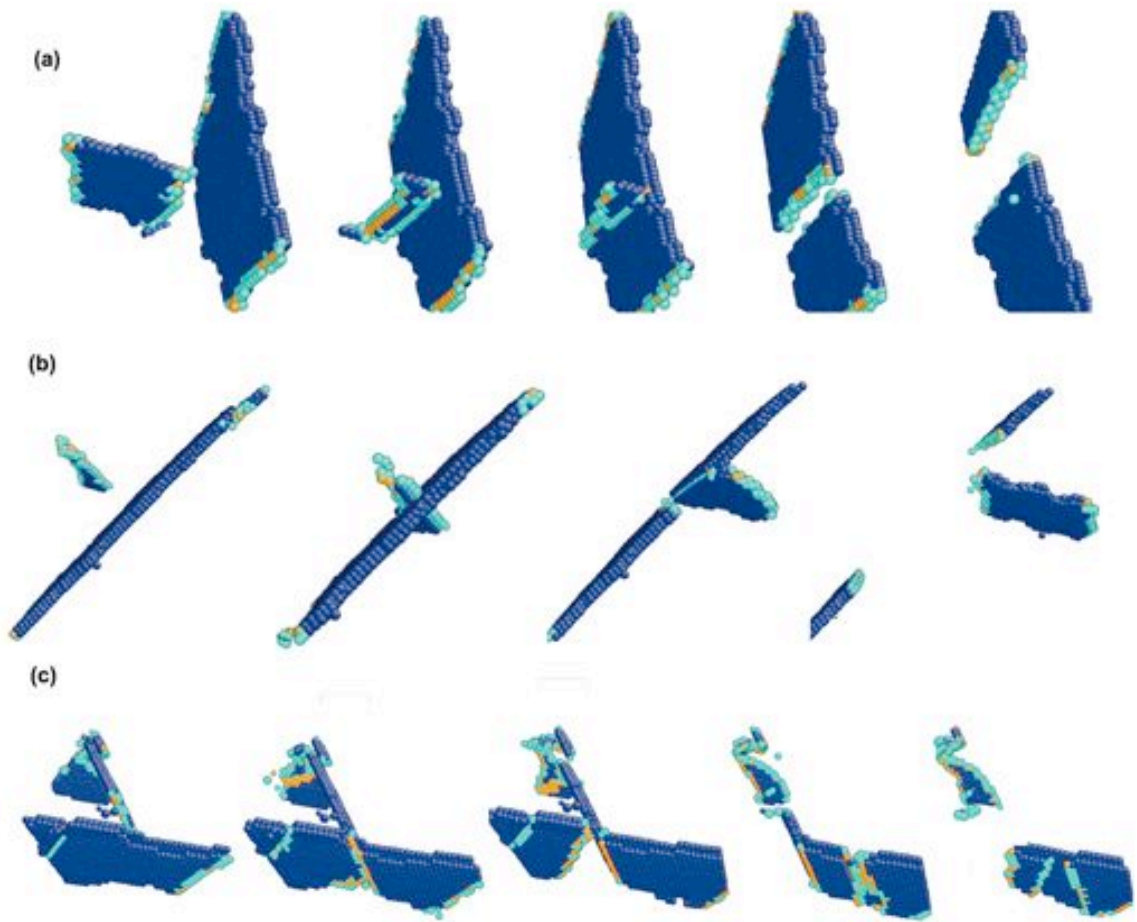

28